\documentclass[conference]{IEEEtran}
\IEEEoverridecommandlockouts
\usepackage{cite}
\usepackage{amsmath,amssymb,amsfonts}
\usepackage{graphicx}
\usepackage{textcomp}
\usepackage{booktabs}
\usepackage{xcolor}
\def\BibTeX{{\rm B\kern-.05em{\sc i\kern-.025em b}\kern-.08em
    T\kern-.1667em\lower.7ex\hbox{E}\kern-.125emX}}
\begin{document}

\title{Simple Trust Metric in a Low-Power Sensor Network}

\author{\IEEEauthorblockN{Svea Wisy}
\IEEEauthorblockA{\textit{Intelligent Systems} \\
\textit{Christian-Albrechts-Universit\"at zu Kiel}\\
Kiel, Germany \\
stu214538@mail.uni-kiel.de}
}

\maketitle

\begin{abstract}
Distributed systems become more and more important to our life. Especially in areas like Smart Home and the Internet of Things (IoT) reliable low-power sensor networks become increasingly important. For ensuring this there are a lot of trust metrics.
In this paper we compare a model of a distributed low-power sensor network including one root node and the corresponding Simple Trust Metric to the requirements from ``Representation of Trust and Reputation in Self-Managed Computing Systems'' \cite{c1}, the Weighted Trust Metric and the Weighted Simple Exponential Smoothing Trust Metric.
\end{abstract}

\begin{IEEEkeywords}
End-to-End Trust, Trust, Metric, Trust Metric, Wireless Sensor Network
\end{IEEEkeywords}

\section{Introduction}
Distributed systems have to handle a lot of changes during their run-time. They especially have to deal with malicious behaviour and failures. For detecting those the authors Kantert et. al. established two requirements for trust metrics in \cite{c1} in the following named ``Reference Article''. They also introduce the Weighted Trust Metric (WTM) and the Weighted Simple Exponential Smoothing Trust Metric (WSES). The same group of researchers gives a low-power sensor network and the Simple Trust Metric in their research article \cite{c2}, in the following as ``End-to-End Trust'', where the Simple Trust Metric is adjusted to the network's task.

Resulting on reading both the articles we wondered if the Simple Trust Metric in End-to-End Trust fulfils the requirements made in the Reference Article and how it differs from the Weighted Trust Metric and the Weighted Exponential Smoothing Trust Metric.

This paper is organised as follows: Section II introduces as well into the definitions of trust and reputation from the Reference Article as into the model of the low-power sensor network from End-to-End Trust. Afterwards, Section III defines the three used trust metrics from the literature. Based on Section II and III, Section IV discusses whether the metric from End-to-End Trust fulfils the requirements from the definition in Section II and how the Simple Trust Metric differs from WTM and WSES. Lastly, Section V concludes the paper with the results. 

\section{Scenario}
To compare the Simple Trust Metric to the requirements from the Reference Article we need to know them as well as we need to know the definition of reputation and trust in general. Second, we need to know the model from End-to-End Trust to have an even-handed comparison between the different trust metrics.

\subsection{Low-power Sensor Network}

The model of End-to-End Trust is a Wireless Sensor Network where each sensor is modelled as one node. Because of the low-power and little memory of the sensors, there is a root node in the network which can store the sensed data and is typically connected to the internet. As defined in the ``IPv6 Routing Protocol for Low-Power and Lossy Networks''\cite{c3} (RPL) the sensed data is sent to the root node, which either stores the data or transmits it via the internet to another node outside the network which can process the sensed data. 

The authors use the option of the RPL, to store the routes used for sending, to periodically send packets from the root back to the nodes. The route is organised in a Destination Oriented Directed Acyclic Graph (DODAG) \cite{c4} which shows the route a packet took from the node to the root. Following this DODAG the authors send packets containing a DODAG Information Object (DIO) from the root back to the nodes.

For improving reliability the authors of End-to-End Trust added a header to the in the RPL defined DIO which contains the Trust Round $t \in \mathbb{N}_{\geq 0}$ and the count of all in the round received packets. The Trust Round is an event-based clock which increases if a DIO is triggered. Due to either non-existent or unsynchronised clocks, it is needed to determine whether a packet was sent before or after the DIO was generated by using this type of clock. So this is where the Trust Rounds come into account. Using the Trust Rounds and the sequence number of a packet, the root node can decide if an arriving packet with a high sequence number indicates a high loss of packets or if it is a packet belonging to the previous Trust Round and is just delivered late. This decision can lead the root node to trigger a DIO if it detects that a lot of packets are missing. Additionally, a DIO is triggered periodically. Triggering a DIO causes the DODAG to change because the nodes now recalculate their best fitting parent node based on their reputation.

\subsection{Reputation}
For managing malicious behaviour or partially failures the authors of the Reference Article define a reputation $\tau \in \mathbb{R}$ for each node n, which is calculated based on a trust metric $\mathfrak{T}(R)$ and the input $R_n$. A good reputation would be a value near $\tau = 1.0$. In this case, the respective node would behave desirably. A reputation value of $\tau = -1.0$ would indicate malicious behaviour or partial failures.

\subsection{Trust Metrics}
For a trust metric $\mathfrak{T}(R_n)$ the authors of the Reference Article use ratings $r \in \mathbb{R}$ where $-1 \leq r \leq 1$. In this case, a rating r is a good rating if $r > 0$, with $r <0$ it is a bad rating, and otherwise it is a neutral rating. Ratings are stored in $R_n$ whose behaviour must be defined when defining a trust metric. Typically $R_n$ is a FIFO-queue such that a trust metric can implement some sort of forgiveness for bad but old ratings. For adding new ratings to $R_n$ trust metrics define a function $\mathfrak{U}(R_n),r$ where the new rating $r$ is added to $R_n$. The result would be $R_{n+1}$ because n increments when adding a new rating.

In the Reference Article, the authors establish two requirements for trust metrics. In the first requirement, they demand that a newly calculated reputation value based on a good rating, either $r_1$ or $r_2$, is better if it bases on the better rating of those unless the reputation already reaches its maximum value. $\mathfrak{R}$ is defined as the interval the values of $r$ can be in. In the Reference Article $\mathfrak{R}$ is in most of their metrics defined as $\mathfrak{R} := [-1,1]$ and accordingly $\mathfrak{R}_+ := (0,1]$ .
\begin{align}
\forall r_1,r_2 \in \mathfrak{R}_+: \mathfrak{T}(\mathfrak{U}(R_n,r_1)) > \mathfrak{T}(\mathfrak{U}(R_n,r_2))\lor \notag\\ 
\mathfrak{T}(\mathfrak{U}(R_n,r_2))=1, \notag \\
r_1 > r_2 \Rightarrow |\mathfrak{R}_+| > 1 \tag{R1} 
\end{align}
In the second requirement, they claim for all positive ratings $r$ that the new reputation is better than the old one as long as the old one didn’t reach the maximum value for reputation yet. $r_+$ is the set of all good ratings and accordingly $r_-$ would be the set of all bad ratings.
\begin{align}
\forall r \in r_+: \mathfrak{T}(\mathfrak{U}(R_n,r)) > \mathfrak{T}(R_n) \lor \mathfrak{T}(R_n) = 1 \tag{R2}
\end{align}

\section{End-to-End Trust And Trust Metrics}
In the End-to-End Trust, the reputation is used to determine whether the current parent is the best one to choose to send sensed data to the root node with less effort. For calculating the delivery rate the nodes use the information they get periodically in the DIO. A parent can only be re-selected when a new Trust Round starts or the parent becomes unavailable. \cite{c5} When a new Trust Round starts the node resets its counter for the sequence numbers and starts sending with the new Trust Round. They calculate the reputation of their parent nodes $n_p$ using the Simple Trust Metric $T_{n,t}$.

\subsection{Simple Trust Metric}
The node calculates the delivery rate $\Xi$ for the parent node $n_p$ in the Trust Round $t$. It gets the amount of successfully delivered packets via the DIO, where the last packets the root received during this Trust Round is given as a value of the modified header. To identify the amount of sent packets the node utilises its current sequence number counter before resetting it. For calculating the delivery rate $\Xi$ the node uses the number of delivered packets and the amount of sent packets. 

If the Trust Round is $t = 0$ the initial trust value of a node $n$ is $0.5$. In case $t > 0$ a new trust value $T_{n,t}$ is calculated by the old trust value $T_{n,t-1}$, the delivery rate $\Xi$ from the last Trust Round and a weight $\alpha \in \mathbb{R}$ with $0 < \alpha < 1$. Usually, the representation of the ratings $R_{t-1}$ is considered as a FIFO-queue, here it is only one floating-point number representing the reputation of the previous round. The stored ratings $r \in \mathbb{R}$ are in the range between 0 and 1, such that $\mathfrak{R} := [0,1]$.
\begin{align}
T \in [0,1]  \\
T_{n,0} := 0.5  \\
\Xi := \frac{delivered}{sent}
\end{align}
\begin{align}
T_{n,t} := \begin{cases}
\alpha \cdot T_{n,t-1}+ (1-\alpha)\cdot \Xi & n=n_p\\
T_{n,t-1} & \text otherwise
\end{cases}
\end{align}

\subsection{Weighted Trust Metric}
In this metric a node gets a new rating $r \in \mathbb{R}$ in a range between $-1$ and $1$, such that $\mathfrak{R}^w := [-1,1]$ The positive ratings are stored in the FIFO-queue $R^w_+$, the negative ratings are represented in the queue $R^w_-$, but there also is a FIFO-queue $R^w_n$ that contains all ratings. Here $n$ is an iterator that increases whenever a new rating is added. $\mathfrak{U}(R^w_n,r) = R_{n+1}$ adds a new rating to the queue. If the maximum length $k$ of $R_n$ is already reached, the authors just drop the oldest rating and add the new rating. For a new reputation value $\tau^w$ the authors of the Reference Article sum up all the ratings and normalise them by the sum of the strength of the ratings. Since they don't define a model, we can not say where the ratings exactly come from.
\begin{align}
\mathfrak{R}^w := [-1,1], r^w \in \mathfrak{R}^w,\mathfrak{R}^w_n \in \mathfrak{R}^{w^k}\\
\tau^w := \mathfrak{T}^w(R^w_n):= \frac{\Sigma_{r \in R^w_n,r >0}r-\Sigma_{r \in R^w_n,r < 0}-r}{\Sigma_{r \in R^w_n,r >0}r+\Sigma_{r \in R^w_n,r < 0}-r} \notag\\
= \frac{\Sigma_{r \in R^w_n}r}{\Sigma_{r \in R^w_n}|r|}
\end{align}

\subsection{Weighted Simple Exponential Smoothing Trust Metric}
Here the rating $r \in \mathbb{R}$ is in the range $\mathfrak{R}^s := [-1,1]$. For this metric, the authors of the Reference Article used Simple Exponential Smoothing (SES)\cite{c6}. This can be thought of as an advanced rolling average with a weight $\alpha \in \mathbb{R}$ and $0 < \alpha < 1$, where a new value for the average can be weighted differently than the old values. In this way, one can implement some sort of forgiveness for partial failure or malicious behaviour. As in the Weighted Trust Metric, the authors store the positive ratings separately from the negative ratings. Due to the rolling average of SES, both values are a floating-point number. In $R^s_n$ the positive reputation value and the negative reputation value are represented as a tuple $(p_1,p_2)$ with $p_1,p_2 \in \mathfrak{R}^s$. The authors also defined a function $\mathfrak{U}(R^s_{n,r}) = R^s_{n+1,r}$, where $n$ is an iterator that increments if a new rating r is added. This function is for adding a new rating to the storage in the way that a positive rating is calculated to the positive reputation value and a negative rating to the negative reputation value of the tuple $(p_1,p_2)$:
\begin{align}
R^s_{n+1} := \mathfrak{U}^s(R^s_n,r)\notag\\
 := \begin{cases}
(p_1 \cdot \alpha + (1- \alpha) \cdot r, p_2\cdot\alpha) & r > 0, (p_1,p_2) \in R^s_n\\
(p_1 \cdot \alpha, p_2 \cdot \alpha - (1-\alpha)\cdot r) & r < 0, (p_1,p_2) \in R^s_n\\
R^s_n & \text otherwise
\end{cases}
\end{align}
Unlike $R^w_n$ in the Weighted Trust Metric, here is no maximum length $k$ of $R^s_n$, because it is no queue but already a tuple of averages.

With this tuple we calculate the new reputation $\tau^s$ by subtracting the negative reputation value $p2$ from the positive reputation value $p1$ and normalise the result with the sum of both to get the reputation for the corresponding node.
\begin{align}
\tau^s:=\mathfrak{T}^s(R^s_n):=\frac{p_1-p_2}{p_1+p_2},(p_1,p_2)\in R^s_n
\end{align}

\section{Discussion}
For the question of fulfilling the requirements we use the requirements from the Reference Article and adjust them to the Simple Trust Metric from End-to-End Trust in the way, they are meant. For comparison, we use the criteria: Range of the ratings, storage of the ratings, the threshold between good and bad ratings, calculation of the reputation value and if the requirements are fulfilled.

\subsection{Requirements on Simple Trust Metric}
To see, if the Simple Trust Metric fulfils the requirements from the Reference Article, we first need some assumptions. We use the definition from IIIA for the delivery rate $\Xi$. Additionally, $n$ is a node and $t$ is the current Trust Round.
\begin{align}
\Xi \text{ is a rating $r^t$} \\
T_{n,t} \text{ calculates a value of reputation}
\end{align}
From the definition of the delivery rate $\Xi = r$ a rating must be positive because there can't be a negative amount of delivered or sent packets. A packet can first be mapped to a new Trust Round after the sending node received the DIO. In that case, the node resets its counter for sequence numbers, but neither the delivered nor the sent amount of packets can become negative. By that, a bad rating can't be a negative number like in the Reference Article. We define $\mathfrak{R^t} := (0,1]$.

A better reference value than $0$ for deciding whether a rating for the node $n$ is either good or bad is the last known reputation value $T_{n,t-1}$ of the node $n$ where $t$ is the current Trust Round. This is caused by the task of the model, which is to deliver the sensed date from the sensors to the node. If the packet loss on a specific route increases in comparison the previous Trust Round, it is slightly a bad event, which should be punished with a bad rating. On contrary, if the packet loss decreases, it is a good occurrence that should be rewarded with a good rating. Because a node can't choose to take part in the network there is no disadvantage in giving a node, that reached the maximum reputation value of 1, a bad or neutral rating.

If a node's last reputation is lower than the new rating, the new rating is good. In the case, the previous reputation value is higher than the new rating, the rating is bad. In the other case, where the last reputation value and the new rating are equal the rating is neutral.

We also define $\mathfrak{U}(R^t_{t-1}) = R^t_t$, where $R^t_t = T_{n,t}$ because there is only one value in the rolling average, such that we don't need any further calculations to get the reputation value.  
\begin{align}
 \forall \Xi \in \mathfrak{R}: \Xi > 0 \\
 \Xi > T_{n,t-1} \Leftarrow \text{$r^t$ is a good rating}\\
 \Xi < T_{n,t-1} \Leftarrow \text{$r^t$ is a bad rating}\\
 \Xi = T_{n,t-1} \Leftarrow \text{$r^t$ is a neutral rating} 
\end{align}
Since the case $T_{n,t-1}=1$ has no good ratings, as a rating can't be higher than 1, we omit this case in the requirements as it is not useful in the Simple Trust Metric. We rebuilt the first requirement by inserting the assumptions:
\begin{align}
[\forall \Xi_1,\Xi_2 > T_{n,t-1}: T_{n,t} \text{ with } \Xi_1 > T_{n,t} \text{ with } \Xi_2,  \notag\\
\Xi_1 > \Xi_2 \Rightarrow |\mathfrak{R}| > 1] 
\end{align}
In the following we also omit the second part of the first requirement $\Xi_1 > \Xi_2 \Rightarrow |\mathfrak{R}| > 1$ because we already know from the definition of $\mathfrak{R}$ that it includes more than one value, such that this part always will be true.
\begin{align}
\forall \Xi_1,\Xi_2 > T_{n,t-1}: T_{n,t} \text{ with } \Xi_1 > T_{n,t} \text{ with } \Xi_2 \\
\Leftrightarrow \alpha \cdot T_{n,t-1}+(1-\alpha)\Xi_1 > \alpha \cdot T_{n,t-1}+(1-\alpha) \Xi_2 \\
\Leftrightarrow (1-\alpha)\Xi_1 > (1-\alpha)\Xi_2\\
\Leftrightarrow \Xi_1 > \Xi_2
\end{align}
According to Equation (18) it is possible to divide by $(1-\alpha)$ without flipping the comparison sign because $0 < \alpha < 1$ \cite{c6} and due to that always $(1-\alpha) > 0$.
As seen in Equation (16) to Equation (19), the Simple Trust Metric fulfils the first requirement from the Reference Article.
Also, the metric fulfils the second requirement (20). We again omit the last part of the requirement:
\begin{align}
[\forall \Xi > T_{n,t-1}: T_{n,t} > T_{n,t-1}]
\end{align}
\begin{align}
T_{n,t} = \alpha \cdot T_{n,t-1}+(1-\alpha)\cdot \Xi \\
 = \alpha \cdot T_{n,t-1}+(1-\alpha)\cdot (T_{n,t-1} +x) \\
 = \alpha\cdot T_{n,t-1}+T_{n,t-1}+x-\alpha T_{n,t-1}-\alpha x \\
 = T_{n,t-1}+x-\alpha x \\
 > T_{n,t-1}+ 0\\
 = T_{n,t-1}
\end{align}
The element $x \in \mathbb{R}_{\geq 0}$ ensures that $\Xi>T_{n,t-1}$ is always true.
By that, the Simple Trust Metric from End-to-End Trust fulfils both requirements from the Reference Article, although the requirements didn't fit without some adjusting to the model. 

This is caused by the differences in the models because in the model of the Reference Article the node itself decides whether to do or not to do a job in the network. In the model of End-to-End Trust, the other nodes decide whether they route their packets using the corresponding node or not. This is decided on the reputation value a node calculates for itself that is calculated by the Simple Trust Metric. So if a node has a bad reputation value calculated for itself, it will at some point decide to take another parent node. Which one the best parent node is, is decided by the Objective Function \cite{c5}.

\subsection{Comparison to Weighted Trust Metric}
After they set up their requirements, the Weighted Trust Metric is the authors first approach in the Reference Article. Its ratings $r^w\in\mathfrak{R}^w$ are in the range between -1 and 1 and the threshold whether a rating is good or bad is the absolute number $0$. In the Simple Trust Metric from End-to-End Trust, the ratings are in the interval (0,1], because $\Xi$ can't become negative due to its definition and the model. The threshold value, on which to decide whether a rating is either good, bad or neutral, isn't fixed but the floating value $T_{n,t-1}$, which is the last reputation value.

Another difference is the different representation $R_n$ of the ratings. WTM is the author's last approach with a FIFO-queue of length $k \in \mathbb{N}$ and the iterator $n$ that increases each time when adding a new rating. The Simple Trust Metric has only one floating-point value for the last reputation value where $T_{n,t} = R^t_t$ for a specific node $n$. One could say, that the Simple Trust Metric also has a FIFO-queue but with $k=1$. In this case adding a new rating $\mathfrak{U}(R^t_{t-1},r^t)$ would calculate a new reputation value $\tau^t$, kicks out the previous value and stores the new value.

Due to the different representations of $R_n$ and the different $\mathfrak{R}$, they differ in the calculation of the reputation. The Weighted Trust Metric calculates the reputation value by the sum of all ratings, normalised by the total amount of all ratings $r^w \in R^w_n$. This way all ratings are weighted the same. A newer rating in this metric affects the reputation in the same way as the old ones do. The Simple Trust Metric uses a value $\alpha$ with $0 < \alpha < 1$ to give a new rating another weight than an old one has. The old value is multiplied by $\alpha$ and the new value by $(1-\alpha)$. For a large value of $\alpha$ one can easily see that the effect of the new value is less than that of the old ones. Reversely a small value of $\alpha$ gives the new value a high significance. 

Both of the metrics fulfil the first requirement from the Reference Article, but the Weighted Trust Metric doesn't fulfil the second requirement in all cases as the authors of the Reference Article show. The Simple Trust Metric fulfils the second requirement as shown in (11) to (26). \\
\\

\begin{table}[htbp]
\centering
\resizebox{0.5\textwidth}{!}{%
\begin{tabular}{ll}
\toprule
Simple Trust Metric & Weighted Trust Metric\\
\midrule
$r^t\in (0,1]$ & $r^w\in [-1,1]$\\
$T_{n,t-1}$ is float & $R^w_{n}$ is FIFO-queue\\
good rating higher than $\Xi$ & good rating higher than 0\\
bad rating lower than $\Xi$ & bad rating lower than 0\\
$\tau^t=T_{n,t}$ & $\tau^w \mathfrak{T}^w(R^w_n)$\\
weighted with $0 < \alpha < 1$ & normalised by $\Sigma_{r \in R^w_n}|r|$\\
fulfils requirement (R1) & fulfils requirement (R1)\\
fulfils requirememnt (R2) & doesn't fulfil requirement (R2)\\
\bottomrule
\\
\end{tabular}}

\caption[table1]{Comparison of Simple Trust Metric and Weighted Trust Metric}
\label{tab: table1}
\end{table}

\subsection{Comparison to WSES Trust Metric}
Versus the Weighted Trust Metric the Weighted Simple Exponential Smoothing Trust Metric (WSES Trust Metric) is a better metric for comparison. Here we have some differences but also many similarities.

Beginning with the ratings it is evident, that they differ the same way as they do in comparison to the Weighted Trust Metric. In the first case, we have a range from 0 to 1 in the other case we have a range from -1 to 1. This is caused by the calculation for the delivery rate from End-to-End Trust. They can not reach a value lower than 0, because it is impossible to send or deliver a negative amount of packets. While the effect of a rating in the WSES Trust Metric is related to the fixed value of 0, the effect of a rating in the Simple Trust Metric is related to the floating value of $T_{n,t-1}$, which is the reputation value of the last Trust Round. 

This is where it gives us the next difference. In contrast to the WSES Trust Metric, the Simple Trust Metric always gets its new ratings with a new Trust Round. The difference is caused by the model the authors used in End-to-End Trust. They need those Trust Rounds for knowing whether a packet was sent before or after a new DIO was generated, which is relevant for the exact calculation of the new rating $\Xi$.

For calculating the new reputation value both the Simple Trust Metric and WSES Trust Metric need the previous reputation value. In both cases, this can be seen as a FIFO-queue with $k=1$ but does not need to be one. While the Simple Trust Metric uses only one floating-point number, the WSES Trust Metric needs a pair of them with the first element for the positive reputation values and the second element for the negative reputation values. That the Simple Trust Metric only needs one value is caused by the fact, that the Simple Trust Metric has no negative ratings and due to that doesn't need to have a storage for them.

When looking at the calculations for the reputation values, one easily sees some similarities as well as some differences. The first similarity is the value $\alpha$, which in both cases is in the range [0,1]. We also see that the first case and the last case of the WSES Trust Metric are nearly the same as in the Simple Trust Metric. The second case of the WSES Trust Metric is not listed for the Simple Trust Metric because it is not relevant. Thinking of the differences in the ratings, we see that the Simple Trust Metric doesn't need to represent negative ratings because there are none. Caused by the same fact, the Simple Trust Metric doesn't need to calculate the second element of a tuple as it also doesn't store it. Caused by this, the Simple Trust Metric requires less memory on each node. This is a nice benefit for the Simple Trust Metric, which is especially useful for its model of a low-power sensor network.

Another difference in the calculation is that the Simple Trust Metric's function $\mathfrak{U}(R^t_t)$ equals its metric $T_{n,t}$ where $t$ is the Trust Round and, $n$ is the corresponding node. Again this is caused by not having any negative ratings because for this metric only one reputation value is calculated. There is no need for normalising the reputation value into one. Other on the WSES Trust Metric, here we need the function $\mathfrak{U}(R^s_n)$, where $n$ is an iterator, to normalise the pair into one reputation value.

Also, the authors of the Reference Article show, that the WSES Trust Metric fulfils their requirements of a Trust Metric. We showed this in Equation (11) to Equation (26) for the Simple Trust Metric.
\\

\begin{table}[htbp]
\centering
\resizebox{0.5\textwidth}{!}{%
\begin{tabular}{ll}
\toprule
Simple Trust Metric & WSES Trust Metric\\
\midrule
$r^t\in (0,1]$ & $r^s\in [-1,1]$\\
$T_{n,t-1}$ is float & $R^s_{n}$ is a tuple of floats\\
good rating higher than $\Xi$ & good rating higher than 0\\
bad rating lower than $\Xi$ & bad rating lower than 0\\
$\tau^t=T_{n,t}$ & $\tau^s \mathfrak{T}^s(R^s_n)$\\
weighted & weighted\\
fulfils requirement (R1) & fulfils requirement (R1)\\
fulfils requirememnt (R2) & fulfils requirement (R2)\\
\bottomrule
\\
\end{tabular}}

\caption[table2]{Comparison of Simple Trust Metric and Weighted Simple Exponential Smoothing Trust Metric}
\label{tab: table2}
\end{table}

\section{Conclusion}
In the first part of the discussion, we showed, that the Simple Trust Metric from End-to-End Trust fulfils the requirements that were defined in the Reference Article. We omitted the last part of the requirements because they can't become true with the conditions. In the second part, we showed the differences between a metric with Simple Exponential Smoothing and a metric with a standard normalisation. In the third part, we experienced how drastic the effect of changing the range in the ratings can be. Especially the difference in the memory is fascinating. Finally, we can say that the authors of End-to-End Trust fulfil all the requirements made in the Reference Article with their Simple Trust Metric. Additionally, their metric has many similarities to the Weighted Simple Exponential Smoothing Trust Metric.

For future work, we think of evaluating more trust metrics on fulfilling the requirements from the Reference Article. Furthermore, we are looking forward to implementing the Simple Trust Metric in a very small network using Mininet.

\end{document}